# III-V Gate-all-around Nanowire MOSFET Process Technology: From 3D to 4D

J. J. Gu,[1] X. W. Wang,[2] J. Shao,[3] A. T. Neal,[1] M. J. Manfra,[3] R. G. Gordon,[2] and P. D. Ye[1]

[1] School of Electrical and Computer Engineering and Birck Nanotechnology Center, Purdue University, West Lafayette, IN 47906, U.S.A.
[2] Department of Chemistry and Chemical Biology, Harvard University, Cambridge, MA 02138, U.S.A.
[3] Department of Physics, Purdue University, West Lafayette, IN, 47906, U.S.A.
Tel: 1-765-494-7611, Fax: 1-765-496-7443, Email: yep@purdue.edu, jjgu@purdue.edu

**Abstract**

In this paper, we have experimentally demonstrated, *for the first time*, III-V 4D transistors with vertically stacked InGaAs nanowire (NW) channels and gate-all-around (GAA) architecture. Novel process technology enabling the transition from 3D to 4D structure has been developed and summarized. The successful fabrication of InGaAs lateral and vertical NW arrays has led to 4× increase in MOSFET drive current. The top-down technology developed in this paper has opened a viable pathway towards future low-power logic and RF transistors with high-density III-V NWs.

**Introduction**

III-V GAA NWFETs, or III-V 3D transistors, have been experimentally demonstrated by a top-down approach, showing excellent scalability down to channel length ($L_{ch}$) of 50nm [1]. However, the $g_m$, SS, and DIBL are greatly limited by the large EOT of the devices [1]. To improve the electrostatic control, we demonstrate 2× reduction in EOT while maintaining the low gate leakage in this work. Furthermore, although lateral (parallel to the wafer surface) integration of NWs has been demonstrated with high drive current per wire, the overall current drivability of the devices is still limited by the pitch of the NWs. To overcome the bottleneck of drive current, *for the first time*, a top-down process technology has been developed to fabricate vertically stacked (normal to the wafer surface) III-V NWFETs, similar to some explored Si NWFETs [2-3]. We call this new type of NW devices *III-V 4D transistors* to distinguish them from III-V 3D transistors [1] which has only one vertical layer and multiple lateral wires. The experimental results, in this paper, show that the drive current per wire pitch ($W_{pitch}$) greatly increases from 3D to 4D structure. This new device concept is very promising for future high-speed low-power logic and RF applications.

**Experiments**

Fig.1 shows the diagram of III-V 3D and 4D transistors with 1×2 and 3×2 NW arrays respectively. Table 1 summarizes the samples investigated in this paper, showing the detailed splits for the NW array, NW size, EOT and the indium content of InGaAs channels. Sample 1 investigates III-V 3D transistors with 5nm $Al_2O_3$ (EOT=2.2nm) gate dielectric and $In_{0.65}Ga_{0.35}As$ incorporation in the NWs to increase gate electrostatic control and boost channel mobility, respectively. The fabrication process for III-V 3D transistors has been developed in Ref. [1]. $Al_2O_3$/WN high-k/metal gate stack was deposited by atomic layer deposition (ALD). The nanowire width ($W_{NW}$) is varied from 25 to 35nm in 5nm steps. Sample 2 demonstrates the first III-V 4D transistors with 10nm $Al_2O_3$ (EOT=4.5nm) and 3×2 or 3×4 NW arrays. Fig. 2 shows the process flow for III-V 4D transistors. Novel process technologies in addition to those developed in [1] are highlighted in red. Fig. 3 shows the schematic diagram of the key steps in the fabrication process of III-V 4D transistors with 3 vertically stacked layers of InGaAs NWs.

The fabrication of III-V 4D transistors started with a 2 inch semi-insulating InP (100) substrate. The following layers were grown sequentially on the InP substrate: a 100nm undoped InAlAs etch stop layer, an 80nm undoped InP sacrificial layer, and then three layers of 30nm $In_{0.53}Ga_{0.47}As$ channel with a 40nm InP layer in between each channel layer (Fig. 3-1). To achieve uniform source/drain doping to contact each of the three vertically stacked channel layers, a two-step Si implantation was performed at an energy of 20keV and 60keV each with a dose of $1\times10^{14}$ cm$^{-2}$. Rapid thermal annealing at 600°C for 15 seconds in $N_2$ was carried out to activate the dopants (Fig. 3-2). To enable the formation of InGaAs/InP fins with a fin height ($H_{Fin}$) of 200nm via reactive ion etching, 10nm ALD $Al_2O_3$ was deposited (Fig. 3-3) and patterned (Fig. 3-4) as a hard mask to replace the electron beam resist [1], providing excellent etching selectivity and eliminating resist redeposition. A new $Cl_2/O_2$ fin etching process was developed replacing the $BCl_3$ based etching [1] to increase the etch rate and improve the sidewall quality (Fig. 3-5). HCl-based solution was used to release the NW arrays patterned along [100] direction (Fig. 3-6). The InAlAs etch stop layer provides a better control of the selective wet etching process. After 10% $(NH_4)_2S$ passivation, the sample was immediately loaded to the ALD reactor for $Al_2O_3$ and WN deposition at 300°C and 385°C respectively (Fig. 3-7). $CF_4/O_2$ based gate etch process was then performed, followed by the electron beam evaporation of Au/Ge/Ni and liftoff process to form the source/drain contacts (Fig. 3-8). After the formation of the source/drain alloy at 350°C, Cr/Au test pads were defined.

Fig. 4(a) shows a top-view SEM image of a III-V 4D transistor with 4 parallel NW stacks. Fig. 4(b) shows the fin structure after $Cl_2/O_2$ dry etching. The InGaAs channel layers, InP sacrificial layers and $Al_2O_3$ etch mask are highlighted. Fig. 4(c)-(d) show the high resolution cross sectional TEM images of InGaAs 3×1 and 3×4 NW arrays, showing that the ALD gate stack was coated around each nanowire. The $W_{NW}$ for layer 1, 2 and 3 is measured to be 20, 60, and 100nm. A better anisotropic dry etch process needs to be developed to have uniform NWs vertically. The $H_{NW}$ for each layer is 30nm defined by MBE. The ALD process of depositing highly conformal WN films for the gate metal is described in Ref [4]. The excellent step coverage of WN was examined on a sample of holes with 0.3 μm in diameter and 11 μm in depth (aspect ratio = 37:1). As shown in Fig. 5, the film thickness was 41 nm and 32 nm in the top and bottom zones, giving 78% step coverage on a structure with an aspect ratio of 37. A superior conformal ALD WN process is critical for the III-V 4D transistor process. All patterns were defined by a Vistec VB-6 UHR electron-beam lithography system. A Keithley 4200 was used to measure MOSFET output characteristics.

### Results and Discussion

Fig. 6-8 show the well-behaved output characteristics, transfer characteristics, and $g_m$-$V_{gs}$ of a III-V 3D transistor (Sample 1) with $L_{ch}$=50nm and $W_{NW}$=25nm. The current is normalized by the total perimeter of the NW, i.e. $W_G$=2×($W_{NW}$+$H_{NW}$)×(Wire Number). The $I_{ON}$ reaches 1.2mA/μm at $V_{ds}$=1V and $V_{gs}$-$V_T$=1.2V, with maximum $g_m$=1.1mS/μm and 1.4mS/μm at $V_{ds}$=0.5V and 1V respectively. Good off-state performance is also achieved with SS=94mV/dec and DIBL=50mV/V. Fig. 9-11 show the DIBL, SS and $V_T$ scaling metrics for Sample 1 compared with Sample 3 published in Ref. [1]. The 2× reduction in EOT greatly suppressed the short channel effects, evident by the dramatic reduction in DIBL, SS and the improvement in $V_T$ roll-off property. DIBL and SS are unchanged over the entire $L_{ch}$ range and show little dependence on $W_{NW}$, indicating an excellent electrostatic control on gate. No significant $V_T$ roll-off is observed for the $L_{ch}$ studied. Fig. 12-13 show $I_{on}$ and $g_m$ scaling metrics for $W_{NW}$=25, 30, and 35nm. Gradual increase in on-state metrics are obtained due to shorter $L_{ch}$. Higher $I_{ON}$ and $g_m$ are obtained on devices with smaller $W_{NW}$ due to the volume inversion effect [5]. Further scaling of $W_{NW}$ down to the sub-10nm regime may require new nanowire thinning techniques.

Fig. 14 shows the output characteristics of a $L_{ch}$=200nm 4D transistor with 3×4 NW array (Sample 2). The $I_{ON}$ reaches 1.35mA/μm at $V_{ds}$=1V and $V_{gs}$=2V normalized by perimeter. The device shows relatively large off-current, which is limited by the leakage current in the top nanowire layer due to the lack of optimization in our implantation process. Maximum $g_m$ of 0.6mS/μm and 0.85mS/μm are obtained at $V_{ds}$=0.5V and 1V respectively, as shown in Fig. 15. The perimeter-normalized $I_{ON}$ and $g_m$ are comparable to those obtained from the best device of sample 3 with the same EOT. This indicates that the new 4D process technology can provide high quality InGaAs NW despite more complicated fabrication. The source/drain resistance ($R_{SD}$) is extracted to be ~600Ω·μm, which can be improved by optimizing the implantation process. Fig. 16 shows the gate leakage current for sample 1, 2, and 3. Sample 2 (4D) shows similarly low gate leakage as Sample 3 (3D), while Sample 1 (3D) shows a slightly increased leakage at a large gate voltage due to the reduction of EOT. Further EOT scaling on both 3D and 4D structures is achievable. Fig. 17 depicts $g_m$-$V_{gs}$ for III-V 4D transistors with 3×2 versus 3×4 NW arrays, showing increased drivability, e.g. $I_{ON}$, by adding NW stacks. To benchmark the overall current drivability for 3D and 4D technologies, $I_{ON}$ and $g_{m,max}$ are normalized by $W_{pitch}$, defined as the maximum $W_{NW}$ of the vertical NW stacks. For III-V 3D transistors, $W_{pitch}$=$W_{NW}$. The $I_{ON}$/$W_{pitch}$ and $g_m$/$W_{pitch}$ are figures of merit which evaluate the drivability per unit width along the plane of the wafer, as shown in Figs. 18-19. Sample 1 shows 2×$I_{ON}$ and 2.5×$g_m$ enhancement over Sample 3 due to mobility enhancement and EOT scaling. Sample 2 shows additional 2×$I_{ON}$ and 1.5×$g_m$ enhancement over Sample 1 due to the introduction of 4D structure despite its larger EOT. Maximum $I_{ON}$/$W_{pitch}$ and $g_m$/$W_{pitch}$ reaches 9mA/μm and 6.2mS/μm for 4D FETs, which can be further improved by reducing the EOT and $R_{SD}$ [6-9].

### Conclusion

We have experimentally demonstrated, for the first time, the fabrication process of III-V 4D transistors using ALD high-k/metal gate stacks. Electrostatic control has been improved by 2× EOT scaling. Record high $I_{ON}$/$W_{pitch}$ and $g_m$/$W_{pitch}$ of 9mA/μm and 6.2mS/μm has been obtained for the III-V 4D transistors, showing 4× improvement over the III-V 3D transistors. The III-V 4D transistor structure is very promising for future high-speed low-power logic and RF applications.

### Acknowledgement

The authors would like to thank X. L. Li, M. S. Lundstrom, D. A. Antoniadis, and J. A. del Alamo for the valuable discussions. This work is supported by the SRC FCRP MSD Center, NSF and AFOSR.

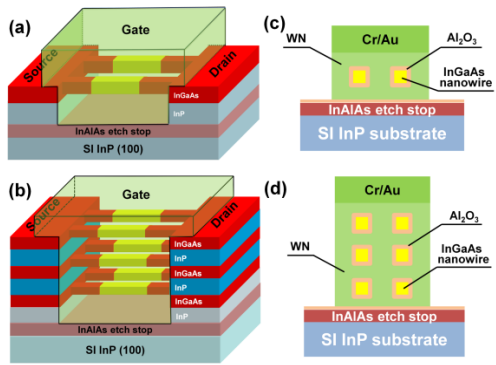

Fig. 1 Schematic diagram of III-V (a) 3D and (b) 4D transistors. (c) and (d) show the cross sectional views of the NWs for (a) and (b), showing 1×2 and 3×2 NW arrays, respectively.

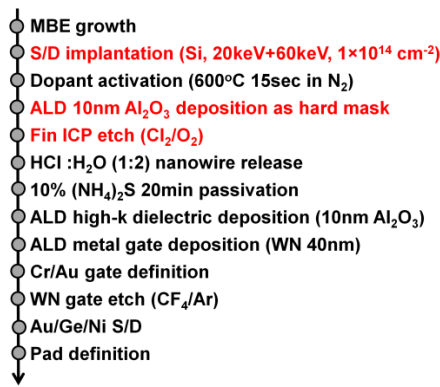

Fig. 2 Fabrication process flow for the III-V 4D transistors with vertical and lateral integrations of InGaAs NWs. Novel process technologies in addition to those developed in Ref. [1] are highlighted in red.

Table 1 NW array (vertical×lateral), NW size and EOT splits for Samples 1 and 2 fabricated in this work, and Sample 3 in Ref. [1].

| | Structure | Nanowire Array (H×W) | $Al_2O_3$ Thickness (nm) | $W_{NW}$ (nm) | $H_{NW}$ (nm) | $In_xGa_{1-x}As$ (x) |
|---|---|---|---|---|---|---|
| Sample 1 (This work) | 3D | 1×4 | 5 | 25 / 30 / 35 | 30 | 65% |
| Sample 2 (This work) | 4D | 3×2 / 3×4 | 10 | 20 (Layer1) / 60 (Layer2) / 100 (Layer3) | 30 | 53% |
| Sample 3 (IEDM11) | 3D | 1×1 / 1×4 / 1×9 / 1×19 | 10 | 30 / 50 | 30 | 53% |

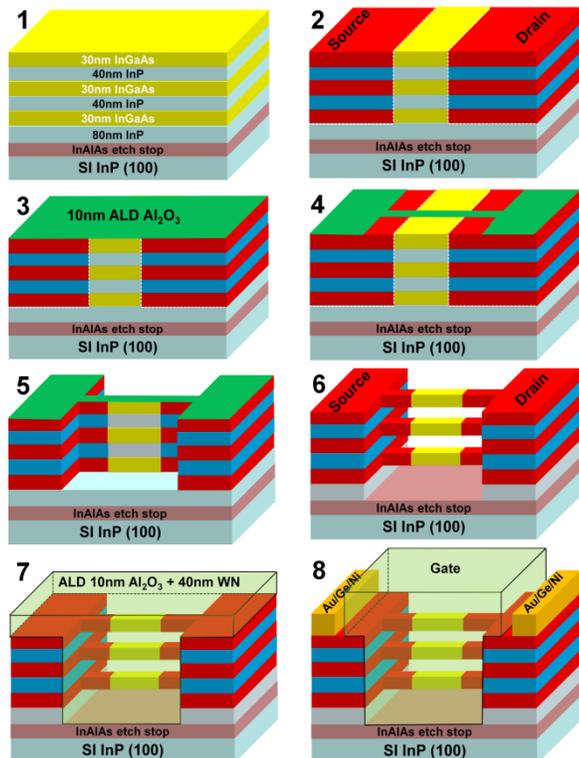

Fig. 3 Schematic diagram of the key process steps in the fabrication of III-V 4D transistors with 3 layers of InGaAs NWs stacked vertically.

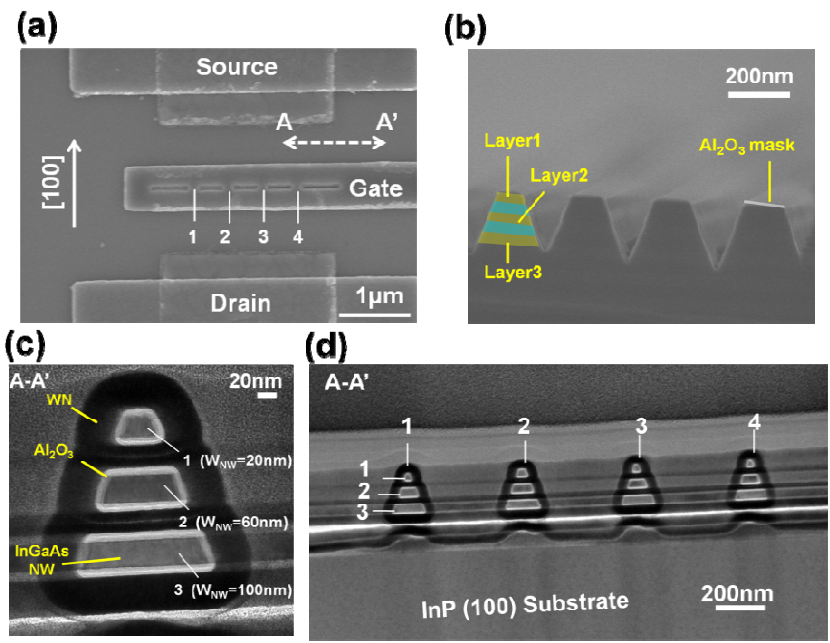

Fig. 4 (a) Top-view SEM image of a III-V 4D transistor with 4 parallel NW stacks. (b) Cross-sectional SEM image of a InGaAs/InP fin test structure fabricated by $Cl_2/O_2$ dry etching with ALD $Al_2O_3$ as a hard mask (c) Cross-sectional TEM image of a 3×1 InGaAs NW stack. The $W_{NW}$ for layer 1, 2 and 3 is measured to be 20, 60, and 100nm. A better anisotropic dry etch process needs to be developed. The $H_{NW}$ for each layer is 30nm defined by MBE. (d) Cross-sectional TEM image of a 3×4 NW array on a real device.

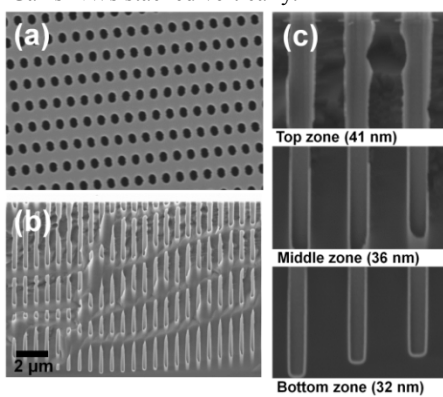

Fig. 5 (a) Top-view and (b)-(c) cross sectional SEM images of an ALD WN coated hole sample with an aspect ratio of 37:1. The film thickness of 41nm, 36nm, and 32nm was obtained in top, middle and bottom zones, indicating 78% step coverage.

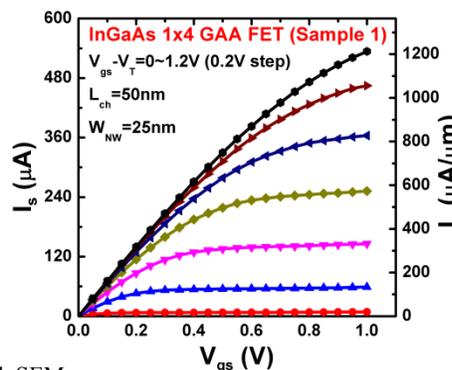

Fig. 6 Output characteristics of a III-V 3D FET with 1×4 NW array (Sample 1). Current is normalized by the total perimeter of the NWs.

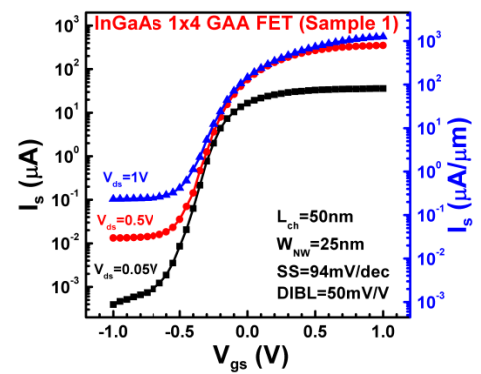

Fig. 7 Transfer characteristics of a III-V 3D FET with 1×4 NW array (Sample 1).

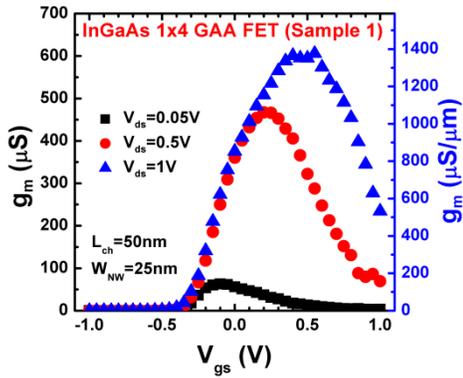

Fig. 8 $g_m$-$V_{gs}$ of a III-V 3D FET with 1×4 NW array (Sample 1).

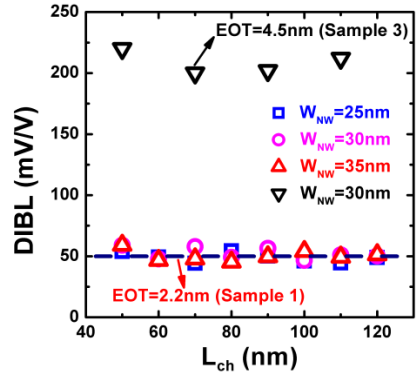

Fig. 9 DIBL scaling metrics for a III-V 3D FET with 1×4 NW array (Sample 1) compared with Sample 3 in Ref. [1].

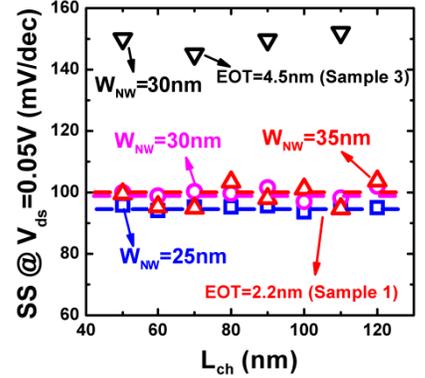

Fig. 10 SS scaling metrics for a III-V 3D FET with 1×4 NW array (Sample 1) compared with Sample 3 in Ref. [1].

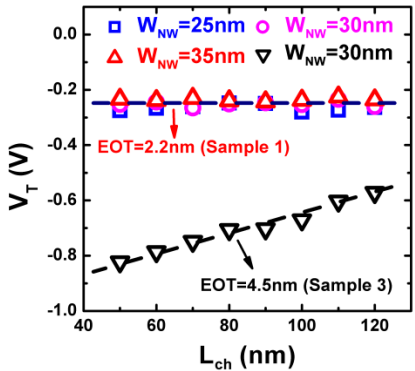

Fig. 11 $V_T$ scaling metrics for a III-V 3D FET with 1×4 NW array (Sample 1) compared with Sample 3 in Ref. [1].

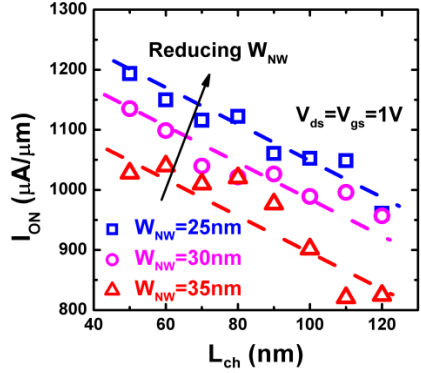

Fig. 12 $I_{ON}$ scaling metrics for III-V 3D FETs with 1×4 NW arrays (Sample 1), showing increased $I_{ON}$ with a reduced $W_{NW}$.

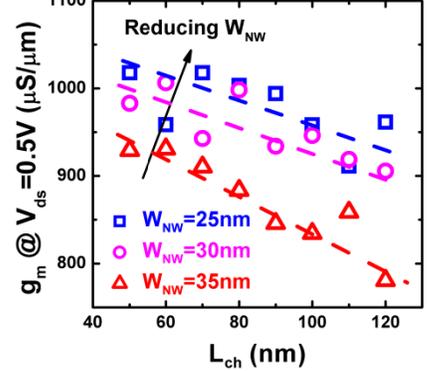

Fig. 13 $g_m$ scaling metrics for III-V 3D FETs with 1×4 NW arrays (Sample 1), showing increased $g_m$ with reduced $W_{NW}$.

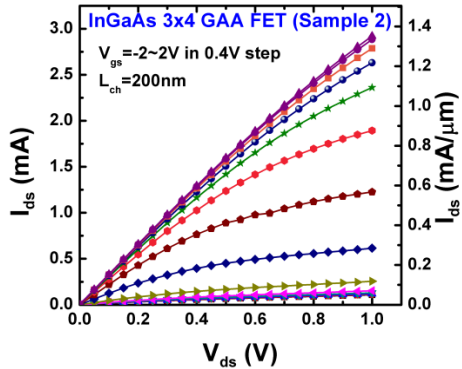

Fig. 14 Output characteristics of a III-V 4D FET with 3×4 NW array (Sample 2).

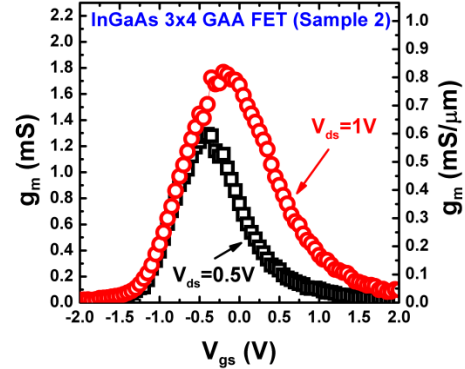

Fig. 15 $g_m$-$V_{gs}$ of a III-V 4D FET with 3×4 NW array (Sample 2).

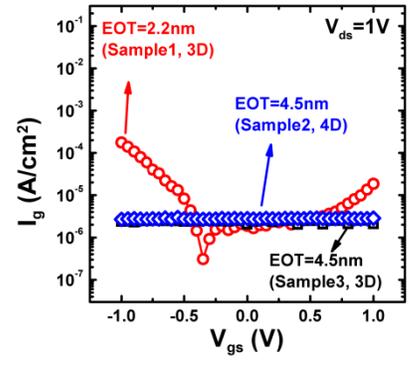

Fig. 16 Gate leakage current density for III-V 3D and 4D FETs (Samples 1, 2, and 3).

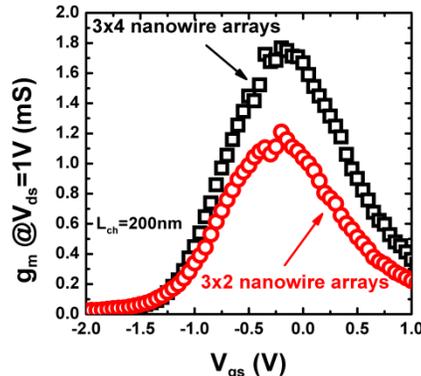

Fig. 17 $g_m$-$V_{gs}$ of III-V 4D FETs with 3×2 and 3×4 NW arrays. The 2× increase in NW number results in ~1.5× increase in maximum $g_m$ due to the variation in NW size, which can be improved by optimizing the fabrication process.

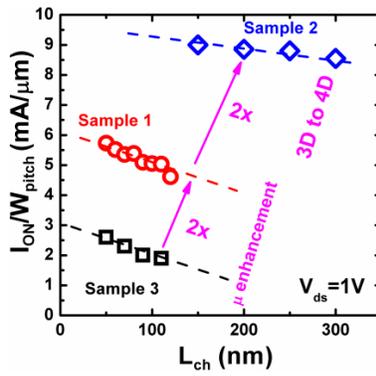

Fig. 18 Benchmarking $I_{ON}$ per $W_{pitch}$ for Samples 1, 2, and 3, indicating the benefit of EOT scaling, mobility enhancement and vertical NW stacking from 3D to 4D integration.

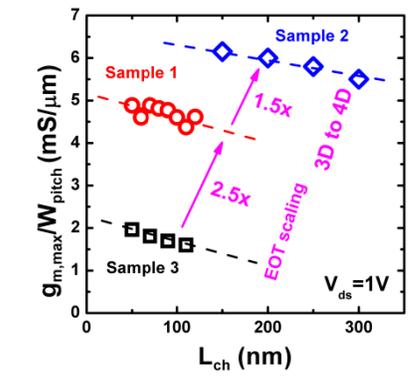

Fig. 19 Benchmarking $g_m$ per $W_{pitch}$ for Samples 1, 2, and 3, indicating the benefit of EOT scaling, mobility enhancement and vertical NW stacking from 3D to 4D integration.